\documentclass{ws-mpla}
\begin{document}

\markboth{Saffari \& Rahvar} {Consistency Condition of Spherically
Symmetric Solutions in $f(R)$ Gravity}

\catchline{}{}{}{}{}

\title{Consistency Condition of Spherically Symmetric Solutions in $f(R)$ Gravity}


\author{Reza Saffari}
\address{Institute for Advanced Studies in Basic Sciences, P.O. Box
45195--1159, Zanjan, Iran, \footnote{rsaffari@iasbs.ac.ir}}

\author{Sohrab Rahvar}%
\address{Department of Physics, Sharif University of Technology,
P.O.Box 11365-–9161, Tehran, Iran,\footnote{rahvar@sharif.edu}}

\maketitle

\pub{Received (Day Month Year)}{Revised (Day Month Year)}

\begin{abstract}
In this work we study the spherical symmetric solutions of $f(R)$
gravity in the metric formalism. We show that for a generic $f(R)$
gravity, the spherical symmetric solution is consistent with the
modified gravity equations except in the case of imposing an extra
condition for the metric.
 \keywords{Cosmology; Dark Energy; Modified
Gravity.}

\end{abstract}
\ccode{95.36.+x, 98.80.Jk, 98.80.Es}

\hspace{.3in}
\newpage

 Recent observations of the Supernova Type Ia and Cosmic
Microwave Background (CMB) radiation indicate that universe is under
positive accelerating expansion ~\cite{wmap3a,spe,per,riess}. One of
the possible solutions to explain the dynamics of universe is
replacing modified gravity models with the Einstein-Hilbert action
\cite{carr04,noj03,noj04,olm,bag,mov,rah}. The simplest class of
this models is using a generic action of $f(R)$ instead of $R$ in
the action. The aim regarding $f(R)$ gravity models is that, these
models can cover all the domains from the cosmological to the solar
system scales. In the cosmological scales some of the models not
only can provide a late time acceleration for the universe but also
predict an inflationary phase for the early universe \cite{sot06}.
In the solar system scales, there is a long list of works in the
spherically symmetric solution of $f(R)$ gravity both in the metric
and the Palatini formalisms
\cite{sss,cli05,cli06,eri,all,nav,chi,olm,gon,sei,soba,sobb,pioneer,mul,capa,capb,capc}.
In the Palatini formalism the solution of vacuum space is a
Schwarzschild-de'Sitter metric with an effective cosmological
constant obtain from the vacuum solution of $f(R)$ gravity.

In the metric formalism recent attempts is using the inverse method
to derive the solar system and galactic scale dynamics. However we
should note that the solution may not satisfy the field equations if
we impose an extra condition on the metric elements
\cite{mul,capa,capb,capc}. Here in this work we revisit the
spherically symmetric solutions of modified gravity in the metric
formalism with more details and present the missed points in the
recent literatures.

We start with a generic form of metric for the spherically symmetric
space:
\begin{equation}
ds^2=-B(r)dt^2+\frac{X(r)}{B(r)}dr^2+r^2(d\theta^2+sin^2\theta
d\phi^2).\label{01}
\end{equation}
For a generic form of Lagrangian as a function of Ricci scalar, the
action is written as:
\begin{equation}
S = \frac{1}{2\kappa}\int d^4x\sqrt{-g}f(R) +S_m.
\label{generalized}
\end{equation}
Varying action with respect to the metric results in the field
equation as:
\begin{equation}
F(R)R_{\mu \nu}-\frac{1}{2}f(R)g_{\mu
\nu}-(\nabla_{\mu}\nabla_{\nu}-g_{\mu \nu}\Box) F(R)=\kappa T_{\mu
\nu}, \label{field}
\end{equation}
where $F=df/dR$ and $\Box\equiv g^{\mu\nu}\nabla_\mu\nabla_\nu$.
From equation (\ref{field}), we take trace and obtain action in
terms of $f$, $F$ and Ricci scalar
\begin{equation}
f(R) = \frac{1}{2}(3\Box F+FR-\kappa T). \label{trace}
\end{equation}
Substituting $f(R)$ from equation (\ref{trace}) in (\ref{field}),
field equation obtain as:
\begin{equation}
R_{\mu\nu}-\frac{1}{4}g_{\mu\nu}R=\frac{\kappa}{F}(T_{\mu\nu}-\frac14g_{\mu\nu}T)
+\frac{1}{F}(\nabla_\mu\nabla_\nu F-\frac{1}{4}g_{\mu\nu} \Box
F).\label{palfe2}
\end{equation}
This equation is diagonal, depends only on $r$. For simplicity in
calculation we rewrite Eq. (\ref{palfe2}) as:
\begin{equation}
K_{[\mu]}=\frac{FR_{\mu\mu}-\nabla_\mu\nabla_\mu F-\kappa
T_{\mu\mu}}{g_{\mu\mu}}, \label{comb}
\end{equation}
where $K_{[\mu]}$ is an index independent parameter. Here the right
hand side of $\mu = t,r,\theta$ terms except $\mu=\phi$, are
independent and the field equation reduces to three independent
equation. On the other hand the constrain of
$K_{[t]}=K_{[r]}=K_{[\theta]}$, reduces the number of independent
equations to two. For the vacuum space $T_{\mu\nu}=0$,
$K_{[t]}-K_{[r]}=0$ results in:
\begin{equation}
\frac{X'}{X}=\frac{2rF''}{2F+rF'},\label{02}
\end{equation}
and from $K_{[t]}-K_{[\theta]}=0$,
\begin{equation}
B''+(\frac{F'}{F}-\frac{1}{2}\frac{X'}{X})B'
-\frac{2}{r}(\frac{F'}{F}-\frac{1}{2}\frac{X'}{X})B
-\frac{2}{r^2}B+\frac{2}{r^2}X=0. \label{03}
\end{equation}
We note that since we are working in vacuum space, the generalized
Bianchi identity provides no more extra constrain on the field
equations.

Now our aim is to extract $B$ and $X$ as the metric elements from
equations (\ref{02}) and (\ref{03}). These two differential
equations contain one more function of $F(r)$. This means that to
have a unique solution for the metric we have to fix either one of
the metric elements (i.e. $X$, $B$) or the action $F$. We study both
cases. Fist let us start with fixing one of the metric elements.

{\bf Fixing metric elements:} Let us assume fixing $X(r)$. In this
case $F(r)$ can be obtained from the equation (\ref{02}) and
substituting this term and $X(r)$ in equation (\ref{03}) we can
obtain $B(r)$. The same procedure can be applied for the case of
fixing $B(r)$. From the metric elements one can extract the Ricci
scalar in terms of $r$. Eliminating $r$ in favor of $R$ in the
expression, $F(r)$ we can obtain the action as a function of Ricci
scalar. This approach is so-called inverse problem, means that
knowing the dynamics of a test particle in the spherically symmetry
space we can demand for a proper action for the gravity
\cite{soba,sobb,pioneer,mul}. As a simple example let us choose
$X=1$. In this case the equation (\ref{02}) reduce to $F''=0$ and
the solution is
\begin{equation}
F(r) = a + br.
\label{result}
\end{equation}
Now substituting this action and $X=1$ in equation (\ref{03})
results in a differential equation for $B$:
\begin{equation}
B''-\frac{2}{r^2}B+\frac{2}{r^2}
+\frac{b}{a+br}(B'-\frac{2}{r}B)=0,\label{sp09}
\end{equation}
The exact solution of this differential equation is:
\begin{eqnarray}
B(r)&=&1-\frac{b}{a}r+(\frac{3}{2}+\ln\big|\frac{b}{a}+\frac{1}{r}\big|)
\frac{b^2}{a^2}r^2+C_1r^2\nonumber\\&~&-C_2\big[\frac{1}{3r}-\frac{b}{2a}
+\frac{b^2}{a^2}r-\frac{b^3}{a^3}r^2\ln\big|\frac{b}{a}+\frac{1}{r}\big|
 \big].\label{bb}
\end{eqnarray}
For the case of $b=0$, equation (\ref{result}) reduces to the
Einstein-Hilbert action and from the equation (\ref{bb}) we recover
the Schwarzschild-de'Sitter metric.

Now we seek the other special solution, choosing $a=0$. In this case
the following solution obtain for the metric from the equation
(\ref{sp09}):
\begin{equation}
B(r) = \frac12 + C_1 r^2+ C_2 r^{-2},
\end{equation}
The corresponding Ricci scalar of this metric is:
\begin{equation}
R = \frac{1}{r^2}- 12C_1.
\end{equation}
Substituting this equation in $F(r) = br$ and integrating it results
in the action in terms of Ricci scalar:
\begin{equation}
f(R) = 2b\sqrt{R + 12C_1}+C_3.
\end{equation}
For the generic case where $a,b\neq 0$, action can only be
calculated numerically.

Depending on our desired dynamics in the spherically symmetric space
we can calculate proper action by this method. For instance one can
use the dynamics of stars in the galaxy to replace the dark matter
with a proper action \cite{rah,soba,pioneer}. For the case of
choosing a complicated function for $X(r)$, the equations (\ref{02})
and (\ref{03}) should be calculated numerically.

{\bf Action fixing:}
 The second approach is fixing action $f(R)$, means
that for a given action, we want to obtain the metric elements, $X$
and $B$. We substitute the action in the equations (\ref{02}) and
(\ref{03}) where the Ricci scalar is expressed in terms of the
metric elements as follows:
\begin{equation}
R=-\frac{1}{X}\left[B''+\frac{4}{r}B'+\frac{2}{r^2}B
-\frac{X'}{X}(\frac{1}{2}B'+\frac{2}{r}B)\right]+\frac{2}{r^2},\label{06}
\end{equation}
We can determine the metric elements by the numerical solution of
the differential equations (\ref{02}) and (\ref{03}). An important
point in this procedure is that putting an extra condition for the
metric elements may cause inconsistent solutions
\cite{mul,capa,capb,capc}. The reason is very simple, adding one
more constrain will decrease the number of parameters to one, while
we have two independent differential equations of (\ref{02}) and
(\ref{03}). For instance let us choose $X=1$ constrain on the metric
elements. From equation (\ref{06}) we can obtain $B(r)$ in terms of
Ricci scalar as follows:
\begin{equation}
B(r)=1+\frac{C_1}{r}+\frac{C_2}{r^2}-\frac{1}{r^2} \int\big[\int
r^2R(r)dr\big]dr,\label{sp011}
\end{equation}
where $R(r)$ can be obtained from $F''(r)=0$. We rewrite this
equation after changing the differentiation to the Ricci scalar as
follows
\begin{equation}
R''f^{(2)}(R)+R'^{2}f^{(3)}(R)=0,\label{sp01}
\end{equation}
where $f^{(n)}(R)\equiv d^nf/dR^n$. For a given $f(R)$, we can
obtain the dependence of the Ricci scalar to the radial distance.
Comparing $B(r)$ in equation (\ref{sp011}) with that of (\ref{bb})
shows that in general those two equations are not identical. In
another word the metric element in equation (\ref{sp011}) is an
inconsistent solution for the spherically symmetric space.

Summarizing this letter, we pointed out that the modified gravity
equations in the form of $f(R)$, reduce to two independent
differential equations in the spherically symmetric space. These
equations contains $X$, $B$ as the metric elements and $F$ as the
derivative of the action. The metric elements as well as the action
can be obtained by fixing one of the them. Fixing $X$ or $B$,
so-called the inverse problem in $f(R)$ gravity means that we can
derive a proper action if we know the dynamics in the spherically
symmetric space. In the second approach one can introduce the action
and calculate the metric elements. Finally we emphasis on
considering extra conditions for the metric, which results
inconsistent solutions for the space. This point is missed in some
of the literatures that try to replace the dark matter with the
modified gravity in the Galactic scales.


\begin{thebibliography}{acc}

\bibitem{wmap3a}
G. Hinshaw {\it et al.}, Astrophys. J. Suppl. {\bf 170}, 288 (2007).

\bibitem{spe}
 D. N. Spergel {\it et al.}, Astrophys. J. Suppl. {\bf 170}, 377
(2007).

\bibitem{per}S.~Perlmutter {\it et al.} (Supernova Cosmology Project
Collaboration), Astrophys.\ J.\ {\bf 517}, 565 (1999).

\bibitem{riess}
A.~G.~Riess {\it et al.} (Supernova Search Team Collaboration),
Astron.\ J.\  {\bf 116}, 1009 (1998).






\bibitem{carr04}
S. M. Carroll, V. Duvvuri, M. Trodden, and M. S. Turner, Phys. Rev.
D {\bf 70}, 043528 (2004).


\bibitem{noj03}
 S.~Nojiri and S.~D.~Odintsov, Phys.\ Rev.\ D {\bf
68}, 123512 (2003).

\bibitem{noj04}
 S.~Nojiri and S.~D.~Odintsov, Gen.\ Rel.\ Grav.\
{\bf 36}, 1765 (2004).

\bibitem{olm}
G.~J.~Olmo, Phys.\ Rev.\ D {\bf 72}, 083505 (2005).

\bibitem{bag}
S. Baghram, M. Farhang, and S. Rahvar, Phys. Rev. D {\bf 75}, 044024
(2007).

\bibitem{mov}
 M. S. Movahed, S. Baghram, and S. Rahvar Phys. Rev. D {\bf
76}, 044008 (2007).

\bibitem{rah}
 S. Rahvar, Y. Sobouti, (accepted in Mod. Phys. Lett. A)
 astro-ph/0704.0680 (2008).

\bibitem{sot06}
T. P. Sotiriou Phys.\ Rev.\ D {\bf 73} 063515 (2006).

\bibitem{sss}
I. Navarro and K. V. Acoleyen, Phys. Lett. B {\bf 622}, 1 (2005).

\bibitem{cli05}
T. Clifton and J. D. Borrow, Phys.Rev. D {\bf 72}, 103005 (2005).

\bibitem{cli06}
T. Clifton, Class. Quant. Grav. {\bf 23} 7445 (2006).

\bibitem{eri}
A. L. Erickcek, T. L. Smith, and M. Kamionkowski, Phys.Rev. D {\bf
74}, 121501 (2006).

\bibitem{all}
G. Allemandi and M. L. Ruggiero, Gen. Rel. Grav {\bf 39}, 1381
(2007).

\bibitem{nav}
I. Navarro and K. V. Acoleyen, JCAP {\bf 0702}, 022 (2007).

\bibitem{chi}
T. Chiba, T. L. Smith and A. L. Erickcek, Phys.Rev. D {\bf 75},
124014 (2007).

\bibitem{olm}
G. J. Olmo, Phys. Rev. Lett. {\bf 98}, 061101 (2007).

\bibitem{gon}
Gonzalo J. Olmo, Phys.Rev. D {\bf 75}, 023511 (2007).

\bibitem{sei}
M. D. Seifert, Phys. Rev. D {\bf 76}, 064002 (2007).


\bibitem{soba}
Y. Sobouti, A\&A {\bf 464}, 921 (2007).

\bibitem{sobb}
R. Saffari and Y. Sobouti, A\&A {\bf 472}, 833 (2007).

\bibitem{pioneer}
R. Saffari and S. Rahvar, Phys. Rev. D {\bf 77}, 104028 (2008).


\bibitem{mul}
T. Multam\"{a}ki and I. Vilja, Phys. Rev. D {\bf 74}, 064022 (2006).

\bibitem{capa}
S. Capozziello, V. F. Cardone and A. Troisi, MNRAS, {\bf 375}, 1423
(2007).
\bibitem{capb}
 S. Capozziello and M. Francaviglia, astro-ph/0706.1146.

\bibitem{capc}
S. Capozziello, A. Stabile and A. Troisi, gr-qc/0709.0891.

\end{thebibliography}
\end{document}